\documentclass[twocolumn,showpacs,superscriptaddress,amsmath,amssymb,aps,pra]{revtex4-1}
\usepackage{graphicx}
\usepackage{CJKutf8}
\usepackage{dcolumn}
\usepackage{amsmath,bm}
\usepackage{epstopdf}
\usepackage[mathlines]{lineno}
\usepackage{color}
\usepackage[colorlinks=true,linkcolor=blue,citecolor=magenta,urlcolor=cyan]{hyperref}

\begin{document}
\title{A spin dephasing mechanism mediated by the interplay between the spin-orbit coupling and the asymmetrical confining potential in semiconductor quantum dot}
\author{Rui\! Li~(\begin{CJK}{UTF8}{gbsn}李睿\end{CJK})}
\email{ruili@ysu.edu.cn}
\affiliation{Key Laboratory for Microstructural Material Physics of Hebei Province, School of Science, Yanshan University, Qinhuangdao 066004, China}
\begin{abstract}
Understanding the spin dephasing mechanism is of fundamental importance in all potential applications of the spin qubit. Here we demonstrate a spin dephasing mechanism in semiconductor quantum dot due to the $1/f$ charge noise. The spin-charge interaction is mediated by the interplay between the spin-orbit coupling and the asymmetrical quantum dot confining potential. The dephasing rate is proportional to both the strength of the spin-orbit coupling and the degree of the asymmetry of the confining potential. For parameters typical of the InSb, InAs, and GaAs quantum dots with a moderate well-height $V_{0}=10$ meV, we find the spin dephasing times are ${\rm T}^{*}_{2}=7$ $\mu$s, $275$ $\mu$s, and $55$ ms, respectively. In particular, the spin dephasing can be enhanced by lowering the well-height. When the well-height is as small as $V_{0}=5$ meV, the spin depahsing times in the InSb, InAs, and GaAs quantum dots are decreased to ${\rm T}^{*}_{2}=0.38$ $\mu$s, $18$ $\mu$s, and $9$ ms, respectively.
\end{abstract}
\date{\today}
\maketitle

\section{Introduction}

It is the existence of the phase coherence that differs a quantum bit (qubit) from a classical bit in information processing, such that a quantum computer potentially can solve certain problems more efficient than a classical computer~\cite{Ladd2010,Buluta2011}. One prerequisite of building a reliable quantum computer is that the building blocks, i.e., the qubits, must have long enough dephasing time~\cite{Nielsen2002}. However, for a realistic experimental qubit candidate such as charge qubit~\cite{Gorman2005,Petersson2010}, spin qubit~\cite{Loss1998,Petta2005,Hanson2007}, and Josephson qubit~\cite{Astafiev2004,You2007,Bylander2011}, the qubit dephasing time is usually severely limited by unexpected and unavoidable environmental noises. Therefore, understanding various qubit dephasing mechanisms is of practical importance to the implementation of quantum computing.

The quantum dot spin qubit has many merits such as the long coherence time~\cite{Veldhorst2014,Veldhorst2015}, the electrical controllability~\cite{Rashba2003,Golovach2006,Tokura2006,Pioro2008,Nowack2007,LiRui2013,Nadj2010,Nadj2012,Schroer2011}, and the convenience for scalability~\cite{Burkard1999,Hu2000,Shulman2012}, so that it is most likely to realize quantum computing in the quantum dot platform. Fluctuating charge field with $1/f$ spectrum has been observed in many quantum nano-systems~\cite{Dutta1981,Weissman1988,Paladino2014}. It also limits the phase coherence time of many qubit candidates~\cite{Astafiev2004,You2007,Bylander2011,Hu2006,Culcer2009,Kha2015}. As recently observed in experiments~\cite{Kawakami2016,Yoneda2018}, the slanting magnetic field in a Si quantum dot mediated a spin-charge interaction, which gave rise to the spin pure dephasing. We are motivated to consider whether the spin-orbit coupling (SOC)~\cite{Bychkov1984}, internally presented in the InSb, InAs, and GaAs quantum dots due to the space-inversion asymmetry, would also mediate a spin dephasing mechanism due to the $1/f$ charge noise?

Quantum dot spin dephasing caused by the charge defects via the combined effects of the SOC and the Coulomb interaction is studied in Ref.~\onlinecite{Bermeister2014}. While the complete quantum theory of the SOC mediated spin dephasing is not well established. Here, let us give a heuristic discussion on how the spin depasing arises in a simple model of the nanowire quantum dot. The Hamiltonian reads~\cite{Levitov2003,Flindt2006,Trif2008,Khomitsky2012,Nowak2013,Romhanyi2015,Ban2015}
\begin{equation}
H=\frac{p^{2}}{2m}+\alpha\sigma^{z}p+\Delta\sigma^{x}+V(x),\label{Eq_model}
\end{equation}
where $m$ is the effective electron mass, $\alpha$ is the Rashba SOC strength~\cite{Bychkov1984}, $\Delta=g\mu_{B}B/2$ is half of the Zeeman splitting, and $V(x)$ is the confining potential. A spin-orbit qubit~\cite{LiRui2013,Nadj2010,Nadj2012} is encoded to the lowest two energy levels (the ground and the first excited states) $\Psi_{\rm e,g}(x)$ of the quantum dot. The qubit couples to the fluctuating charge field $\textbf{E}$ via the electric-dipole interaction $eE_{x}x$~\cite{Scully1999}. The difference between $\langle\Psi_{\rm e}|x|\Psi_{\rm e}\rangle$ and $\langle\Psi_{\rm g}|x|\Psi_{\rm g}\rangle$ leads to a longitudinal interaction between the qubit and the noise, which gives rise to the qubit pure dephasing. While $\langle\Psi_{\rm e}|x|\Psi_{\rm g}\rangle$ leads to a transverse interaction between the qubit and the noise, which gives rise to the possible qubit relaxation. The necessary condition for the qubit phase noise is $\langle\Psi_{\rm e}|x|\Psi_{\rm e}\rangle\neq\langle\Psi_{\rm g}|x|\Psi_{\rm g}\rangle$. However, if the confining potential is symmetrical  $V(x)=V(-x)$, the model (\ref{Eq_model}) has a $Z_{2}$ symmetry $[\sigma^{x}\mathcal{P},H]=0$~\cite{Braak2011,Xie2014,LiRui2018_SR}, where $\mathcal{P}$ is the parity. The $Z_{2}$ symmetry directly leads to $\langle\Psi_{\rm e (g)}|x|\Psi_{\rm e (g)}\rangle=0$. Therefore, the qubit phase noise in our model can arise only when $V(x)$ is an asymmetrical potential.

In this paper, we have formulated a theory of  the SOC mediated spin pure dephasing based on an exactly solvable model of the nanowire quantum dot. We demonstrate the interplay between the SOC and the asymmetrical confining potential mediates a spin-charge interaction, that gives rise to the spin pure dephasing. Both the SOC and the asymmetry of the confining potential are indispensable in this dephasing mechanism. The larger of the SOC in the material, the stronger of the spin depasing. Likewise, the larger of the degree of the asymmetry of the confining potential, the stronger of the spin dephasing. Also, the spin dephasing can be enhanced when we lower the height of the quantum dot confining potential.

\section{The model}

\begin{figure}
\centering
\includegraphics{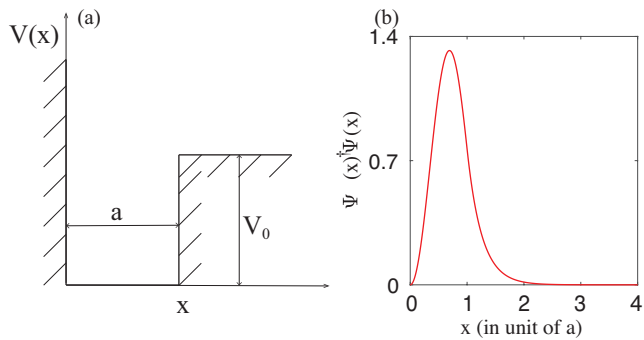}
\caption{\label{Fig_model}(a) The half infinite square well used to model the asymmetrical confining potential of a nanowire quantum dot. The well has both a width $a$ and a height $V_{0}$. (b) The probability density distribution of the ground state in a InSb quantum dot. }
\end{figure}

\begin{table}
\centering
\caption{\label{tab}The parameters of the InSb, InAs, and GaAs quantum dots used in our calculations (Refs.~\onlinecite{Sousa2003,Winkler2003}).}
\begin{ruledtabular}
\begin{tabular}{ccccccc}
~&$m/m_{0}$\footnote{$m_{0}$ is the free electron mass}&$\alpha$~(eV \AA)&$g$&$B_{0}$~(T)&$a$~(nm)&$V_{0}$~(meV)\\
InSb&$0.0136$&$1.05$&$50.6$&$0.05$&$50$&$10$\\
InAs&$0.0239$&$0.23$&$15$&$0.1$&$50$&$10$\\
GaAs&0.067&0.01&0.44&7.5&$50$&$10$
\end{tabular}
\end{ruledtabular}
\end{table}

Here we are interested in a 1D model of the nanowire quantum dot with both asymmetrical confining potential and nontrivial Rashba SOC. The explicit Hamiltonian under consideration is given by Eq.~(\ref{Eq_model}),  and the asymmetrical confining potential is modeled by the following half infinite square well [see Fig.~\ref{Fig_model}(a)]
\begin{equation}
V(x)=\left\{
\begin{array}{cc}
\infty,  & x<0,    \\
 0, & 0<x<a ,  \\
 V_{0}, & a<x,
\end{array}
\right.\label{Eq_well}
\end{equation}
where $V_{0}$ and $a$ are the height and width of the well, respectively. The confining potential has such a regular shape that the bound states in the well are expected to be exactly solvable~\cite{LiRui2018_SR,Bulgakov2001,Tsitsishvili2004,LiRui2018_PRB}. The lowest two energy levels in the quantum dot are used to encode a qubit. In the presence of the nontrivial Rashba SOC, the spin operator in Hamiltonian~(\ref{Eq_model}) is no longer a good quantum number, such that the qubit defined in our model is actually a spin-orbit qubit~\cite{LiRui2013,Nadj2010,Nadj2012}. In our following considerations, the quantum states span the qubit Hilbert space are marked by the pseudo spin states: $\Psi_{\rm e}(x)\equiv|\!\Uparrow\rangle$ and $\Psi_{\rm g}(x)\equiv|\!\Downarrow\rangle$. In contrast to the pure spin qubit, the spin-orbit  qubit has the advantage of being electrically manipulable~\cite{Rashba2003,Golovach2006,Tokura2006,Pioro2008,LiRui2013,Nowack2007,Nadj2010,Nadj2012}.

The boundary condition is used to determine the energy spectrum and the corresponding eigenfunctions of a quantum system. For the square well (\ref{Eq_well}) we are considering, the boundary condition explicitly reads~\cite{LiRui2018_SR}
\begin{equation}
\Psi(0)=0,~\Psi(a+0)=\Psi(a-0),~\Psi'(a+0)=\Psi'(a-0),\label{Eq_boundary}
\end{equation}
where $\Psi(x)$ is the eigenfunction and $\Psi'(x)$ is its first derivative. It should be noted that the eigenfunction $\Psi(x)=[\Psi_{1}(x),\Psi_{2}(x)]^{\rm T}$ here has two components due to the spin degree of freedom. Hence, the boundary condition (\ref{Eq_boundary}) actually contains six independent sub-equations.

Let us say a few words on the model we are considering. First, although our model is very simple, we believe that this model captures the main physics of the SOC mediated spin dephasing in an asymmetrical quantum dot. Second, we expect that the physics (at least qualitatively) in a more realistic 2D quantum dot would be similar to that in our exactly solvable 1D model. Third, as far as we know, there is no exact solution for a 2D quantum dot with both asymmetrical confining potential and non trivial SOC. Therefore, investigating a simple exactly solvable quantum dot model no doubt gives the first step for understanding the relevant properties in a more complicated and more realistic quantum dot.

In this paper, we mainly study three quantum dot materials, i.e., the InSb, InAs, and GaAs, all of which are of current research interest~\cite{Sousa2003,Winkler2003}. The InSb has the largest SOC, the InAs has a relative large SOC, and the GaAs has the smallest SOC. In our following calculations, unless otherwise stated, all the parameters are taken from Table~\ref{tab}.

\section{The qubit Hilbert space structure}

\begin{figure}
\centering
\includegraphics[width=8.5cm]{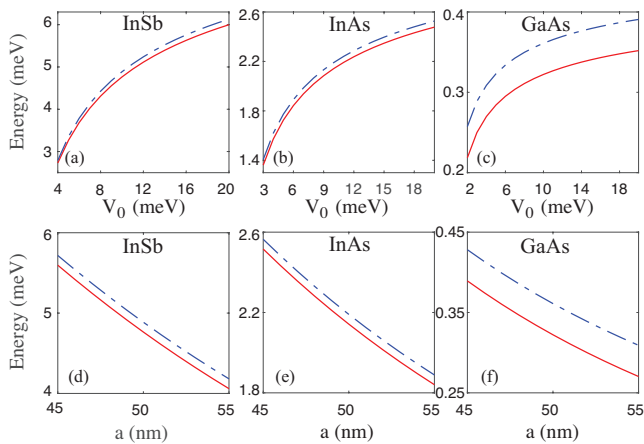}
\caption{\label{Fig_energyspectrum}(a)-(c) The lowest two energy levels as a function of the well-height $V_{0}$. The results in the InSb (a), InAs (b), and GaAs (c) quantum dots. (d)-(f) The lowest two energy levels as a function of the well-width $a$. The results in the InSb (d), InAs (e), and GaAs (f) quantum dots.}
\end{figure}

We first solve the spectrum and the wave functions for the bulk Hamiltonian $H_{\rm b}=\frac{p^{2}}{2m}+\alpha\sigma^{z}p+\Delta\sigma^{x}$~\cite{LiRui2018_SR,LiRui2018_PRB}. Then the eigenfunction of Hamiltonian (\ref{Eq_model}) can be written as a linear combination of all the degenerate bulk wave functions~\cite{Bulgakov2001,Tsitsishvili2004,LiRui2018_PRB}. Inside the well, the eigenfunction can be expanded using both the plane-wave and exponential-function solutions. Outside the well, the eigenfunction can be expanded using either the combined plane-wave and exponential-function solutions or the exponential-function solutions. Imposing the boundary condition (\ref{Eq_boundary}) on the expanded eigenfunction, we obtain a series of transcendental equations with respect to the energy region [see appendices~\ref{appendix_a}, \ref{appendix_b}, and \ref{appendix_c}]. The solutions of these transcendental equations give us the total energy spectrum of the quantum dot. Once the spectrum is obtained, the corresponding eigenfunctions are also known. A typical probability density distribution of the ground state in a InSb quantum dot is given in Fig.~\ref{Fig_model}(b).

In Figs.~\ref{Fig_energyspectrum}(a)-(c), we show the lowest two energy levels as a function of the well-height $V_{0}$ in the InSb, InAs, and GaAs quantum dots, respectively. As can be seen from the figures, with the decease of the well-height $V_{0}$, the energies of the corresponding quantum states become smaller, i.e., more closer to the well-portal, and the qubit level splitting becomes smaller too. This phenomenon has been observed previously, the spin-orbit effect in the quantum dot can be enhanced by lowering the height of the confining potential~\cite{LiRui2018_SR}. We can understand as follows. The quantum dot spin-orbit effect can be roughly characterized by the parameter $\langle\!\langle\,x\rangle\!\rangle/x_{\rm so}$~\cite{LiRui2013}, where $\langle\!\langle\,x\rangle\!\rangle$ is the half-width of the quantum dot wave function [see Fig.~\ref{Fig_model}(b)] and $x_{\rm so}=\hbar/(m\alpha)$ is the spin-orbit length. Obviously, when we lower the well-height $V_{0}$, the wave function is more delocalized, hence $\langle\!\langle\,x\rangle\!\rangle$ becomes larger. It should be noted that, the well-height $V_{0}$ in our model can not be arbitrary small if we want at least two bound states presented in the well.

In Figs.~\ref{Fig_energyspectrum}(d)-(f), we also show the lowest two energy levels as a function of the well-width $a$. There are no obvious changes for the qubit level splitting when the well-width $a$ is varied in the region under consideration. However, with the decrease of the well-width $a$, the energies of the corresponding quantum states becomes larger, i.e., more closer to the well-portal. Likewise, the well-width $a$ also can not be arbitrary small if we want to maintain at least two bound states in the well.

The spin-orbit qubit can couple to the charge noise via the electric-dipole interaction $eE_{x}x$~\cite{Scully1999}, where $E_{x}$ is the $x$ component of the fluctuating charge field. Hence, we need to determine the form of the electric-dipole operator $x$ in the qubit Hilbert space. The phase noise of the qubit arises when the average values of $x$ between the first excited state $x_{\rm e}\equiv\langle\Psi_{\rm e}|x|\Psi_{\rm e}\rangle$ and the ground state $x_{\rm g}\equiv\langle\Psi_{\rm g}|x|\Psi_{\rm g}\rangle$ are different. Since the exact eigenfunctions in the quantum dot are already obtained [see e.g., Fig.~\ref{Fig_model}(b)], these two quantities $x_{\rm e}$ and $x_{\rm g}$ are easy to evaluate. 

\begin{figure}
\centering
\includegraphics[width=8.5cm]{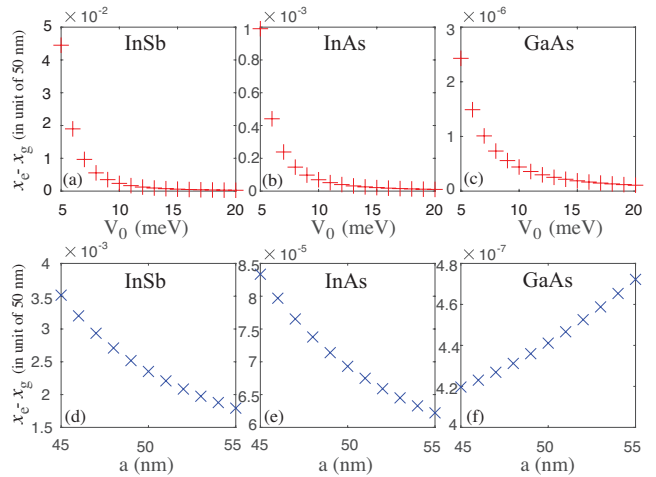}
\caption{\label{Fig_operator}(a)-(c) The difference $x_{\rm e}-x_{\rm g}$ as a function of the well-height $V_{0}$. The results in the InSb (a), InAs (b), and GaAs (c) quantum dots. (d)-(f) The difference $x_{\rm e}-x_{\rm g}$ as a function of the well-width $a$. The results in the InSb (d), InAs (e), and GaAs (f) quantum dots.}
\end{figure}

In Figs.~\ref{Fig_operator}(a)-(c), we show the difference of the averages $x_{\rm e}-x_{\rm g}$ as a function of the well-height $V_{0}$ in the InSb, InAs, and GaAs quantum dots, respectively. In consistence with the $V_{0}$ dependence of the energy spectrum, here with the decrease of the well height $V_{0}$, the difference of the averages $x_{\rm e}-x_{\rm g}$ becomes larger, i.e., the spin-orbit effect becomes stronger. In Figs.~\ref{Fig_operator}(d)-(f), we also show the difference of the averages $x_{\rm e}-x_{\rm g}$ as a function of the well-width $a$. For materials with both strong SOC and relative small effective electron mass such as InSb and InAs, with the decrease of the well-width $a$, the difference $x_{\rm e}-x_{\rm g}$ becomes larger. While for the material with weak SOC and relative large effective electron mass, e.g., GaAs, with the decrease of the well-width $a$, the difference $x_{\rm e}-x_{\rm g}$ becomes smaller instead. Actually, if we continue to reduce the GaAs quantum dot size $a$ to smaller value such as $30$ nm, after a critical value $a_{\rm c}$, the difference $x_{\rm e}-x_{\rm g}$ also increases with the decrease of $a$ until the two bound states are repelled out of the well.

\section{The spin pure dephasing}

\begin{figure}
\includegraphics{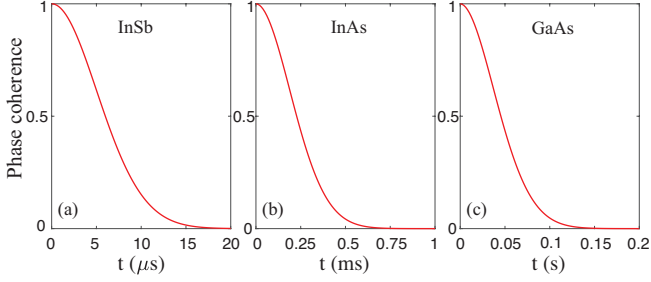}
\caption{\label{Fig_phasecoherenc}The phase coherence $\left|\rho_{\Uparrow\Downarrow}(t)/\rho_{\Uparrow\Downarrow}(0)\right|=e^{-\Gamma_{\rm ph}(t)}$ of the spin qubit as a function of time $t$. (a) In the InSb quantum dot. (b) In the InAs quantum dot. (c) In the GaAs quantum dot.}
\end{figure}

In III-V semiconductor quantum dot, the spin dephasing mechanism caused by the surrounding magnetic noise is well established~\cite{Yao2006,Witzel2006,Cywinski2009}. It is the magnetic dipole interactions between the lattice nuclear spins produce a fluctuating hyperfine field to the electron spin. $1/f$ charge noise universally exists in many quantum nano-structures~\cite{Dutta1981,Weissman1988,Paladino2014}, and it has also been observed in many quantum dot experiments~\cite{Yoneda2018,Jung2004,Kuhlmann2013,Chan2018}, hence it is desirable to examine whether there exists charge noise induced spin dephasing in spin-orbit coupled quantum dot. In particular, in a recent InSb quantum dot experiment, the spin dephasing induced by the $1/f$ charge noise can not be ruled out~\cite{Berg2013}. The physical origin of the $1/f$ charge fluctuation spectrum is still not very clear~\cite{Paladino2014}, here we just assume the charge noise has a spectrum function $\propto1/\omega$.

The spin-orbit qubit in a semiconductor quantum dot can couple to the charge field via the electric-dipole interaction. The total Hamiltonian describing the qubit-noise interaction reads
\begin{equation}
H_{\rm tot}=H+ex\cos\Theta\sum_{k}\Xi_{k}(b_{k}+b^{\dagger}_{k})+\sum_{k}\hbar\omega_{k}b^{\dagger}_{k}b_{k},
\end{equation}
where we have written the fluctuating charge field as ${\bf E}=\sum_{k}\Xi_{k}\vec{e}_{k}(b_{k}+b^{\dagger}_{k})$~\cite{Scully1999}, with $\Xi_{k}$ being the charge field in the wavevector space and $\vec{e}_{k}$ being the direction of the charge field, and $\Theta$ is the angle between $\vec{e}_{k}$ and the axis of the nanowire $\vec{x}$. In our following calculations, we have averaged over all possible angle $\Theta$ for the obtained physical quantities, e.g., $\langle\Gamma(t)\rangle_{\Theta}=\int^{2\pi}_{0}\Gamma(t)d\Theta/2\pi$.

When we focus only on the qubit Hilbert subspace,
the total Hamiltonian can be reduced to (only phase noise is taken into account)
\begin{eqnarray}
H_{\rm tot}&=&\frac{E_{\rm e}-E_{\rm g}}{2}\tau^{z}+\sum_{k}\hbar\omega_{k}b^{\dagger}_{k}b_{k}+\nonumber\\
&&\sum_{k}\left(\frac{x_{\rm e}+x_{\rm g}}{2}+\frac{x_{\rm e}-x_{\rm g}}{2}\tau^{z}\right)e\Xi_{k}(b_{k}+b^{\dagger}_{k})\cos\Theta,\nonumber\\
\end{eqnarray}
where $E_{\rm e,g}$ are the energies of the first excited state $|\!\!\Uparrow\rangle$ and the ground state $|\!\!\Downarrow\rangle$, respectively, the Pauli $z$ matrix reads $\tau^{z}=|\!\!\Uparrow\rangle\langle\Uparrow\!\!|-|\!\!\Downarrow\rangle\langle\Downarrow\!\!|$, and we have also used the completeness relation $|\!\!\Uparrow\rangle\langle\Uparrow\!\!|+|\!\!\Downarrow\rangle\langle\Downarrow\!\!|=1$.
Obviously, if $x_{\rm e}=x_{\rm g}$, the spin-orbit qubit can not longitudinally couple to the charge noise. From this viewpoint, it is the difference of the average values of the electric-dipole operator $x_{\rm e}-x_{\rm g}$, which originates from the interplay between the SOC and the asymmetrical confining potential, gives rise to the pure dephasing of the spin-orbit qubit in semiconductor quantum dot.

The model we derived is very similar to the spin-boson model~\cite{Palma1996,Duan1998,Uhrig2007}. A simple analysis shows that the qubit dephasing of this model is also exactly solvable. If we model the phase coherence as the off-diagonal element of the qubit density matrix $\left|\rho_{\Uparrow\Downarrow}(t)/\rho_{\Uparrow\Downarrow}(0)\right|={\rm exp}\left[-\Gamma_{\rm ph}(t)\right]$, the dephasing rate can be written as (for details see appendix~\ref{appendix_d})
\begin{equation}
\Gamma_{\rm ph}(t)=\frac{(x_{\rm e}-x_{\rm g})^{2}}{2a^{2}}\int^{\omega_{\rm max}}_{\omega_{\rm min}}d\omega\,S(\omega)\frac{\sin^{2}(\omega\,t/2)}{(\omega/2)^{2}},\label{Eq_dephasing}
\end{equation}
where the spectrum function is defined as
\begin{equation}
S(\omega)=\sum_{k}\frac{e^{2}\Xi^{2}_{k}a^{2}k_{B}T}{\hbar^{3}\omega}\delta(\omega-\omega_{k})\equiv\frac{A^{2}_{a,T}}{\omega},
\end{equation}
with $A^{2}_{a,T}$ being a parameter characterizing the strength of the charge noise~\cite{LiRui2018_arXiv}. Here $\omega_{\rm min}$ and $\omega_{\rm max}$ are the lower and the upper bounds of the charge noise spectrum~\cite{Schriefl2006}. Also, we have written the Bose occupation number as $n(\omega)\approx\,k_{B}T/\hbar\omega$ for all the low frequency $1/f$ charge noise mode. In consistence with our previous investigation~\cite{LiRui2018_arXiv}, here we choose the spectrum strength $A_{a=50{\rm nm}, T=100 {\rm mK}}=20$ MHz, and the other parameters of the noise are taken from experiment~\cite{Yoneda2018}, e.g., the lower noise bound $\omega_{\rm min}\approx10^{-2}$ Hz, the upper noise bound $\omega_{\rm max}\approx5\times10^{5}$ Hz, and the typical experimental temperature $T=100$ mK. It is instructive to see for the time scale $t<1/\omega_{\rm max}=2~\mu$s, we can write the dephasing rate as~\cite{LiRui2018_arXiv}
\begin{equation}
\Gamma_{\rm ph}(t)=A^{2}_{a,T}t^{2}\frac{(x_{\rm e}-x_{\rm g})^{2}}{2a^{2}}\ln\frac{\omega_{\rm max}}{\omega_{\rm min}}.\label{eq_gaussdecay}
\end{equation}
Thus, the qubit dephasing at short time must be a Gauss decay.

In Fig.~\ref{Fig_phasecoherenc}, we show the qubit phase coherence as a function of time $t$. For parameters typical of the InSb quantum dot, because of the large SOC, the qubit dephasing time is about ${\rm T}^{*}_{2}=7$ $\mu$s. For a InAs quantum dot, the SOC is still relative large, we find the qubit dephasing time is about ${\rm T}^{*}_{2}=275$ $\mu$s. For a GaAs quantum dot, because of the very weak SOC, the qubit dephasing time is about ${\rm T}^{*}_{2}=55$ ms.

\begin{figure}
\centering
\includegraphics{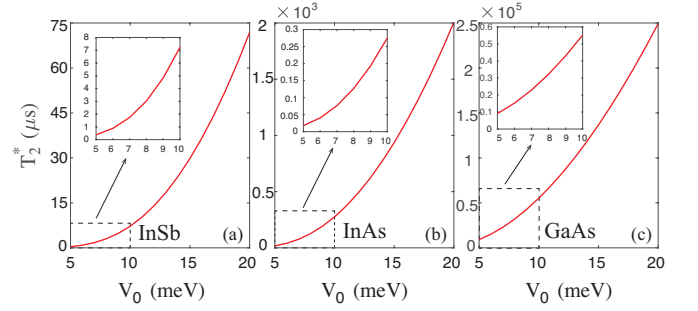}
\caption{\label{Fig_T2vsV0}The dephasing time ${\rm T}^{*}_{2}$ as a function of the well height $V_{0}$. (a) In the InSb quantum dot. (b) In the InAs quantum dot. (c) In the GaAs quantum dot.}
\end{figure}

\begin{figure}
\centering
\includegraphics{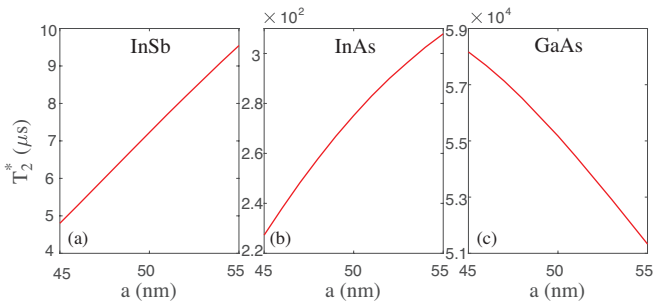}
\caption{\label{Fig_T2vsa}The dephasing time ${\rm T}^{*}_{2}$ as a function of the well width $a$. (a) In the InSb quantum dot. (b) In the InAs quantum dot. (c) In the GaAs quantum dot.}
\end{figure}

In Fig.~\ref{Fig_T2vsV0}, we show the qubit dephasing time ${\rm T}^{*}_{2}$ as a function of the quantum dot well-height. Note that ${\rm T}^{*}_{2}$ is solved from $\Gamma_{\rm ph}({\rm T}^{*}_{2})=1$ defined in Eq.~(\ref{Eq_dephasing}). As expected, the spin-orbit effect in the quantum dot can be enhanced by lowering the well-height~\cite{LiRui2018_SR}, such that the qubit dephasing time ${\rm T}^{*}_{2}$ becomes smaller when we reduce the well-height $V_{0}$. When the well-height is as small as $V_{0}=5$ meV, the dephasing time is about ${\rm T}^{*}_{2}=0.38$ $\mu$s [see Fig.~\ref{Fig_T2vsV0}(a)], $18$ $\mu$s [see Fig.~\ref{Fig_T2vsV0}(b)], and $9$ ms [see Fig.~\ref{Fig_T2vsV0}(c)] in a InSb, InAs, and GaAs quantum dots, respectively. The magnitude of $V_{0}$ reflects the degree of the asymmetry of the confining potential. The larger of the asymmetry of the well, the stronger of the qubit dephasing. Note that in the GaAs quantum dot the spin dephasing time is in the microsecond region~\cite{Petta2005,Koppens2008,Bluhm2011}, in the InAs quantum dot a ${\rm T}_{2}=50$ ns is reported in Ref.~\cite{Nadj2010}, and in the InSb quantum dot a ${\rm T}_{2}=34$ ns is reported in Ref.~\cite{Berg2013}. Because of the large SOC, the $1/f$ charge noise induced dephasing is most likely to be observed in the InSb quantum dot.

In Fig.~\ref{Fig_T2vsa}, we show the qubit dephasing time ${\rm T}^{*}_{2}$ as a function of the quantum dot size $a$. As can be seen from the figure, in the InSb and InAs quantum dots, because the qubit energy levels are shallow energy levels in the well, i.e., close to the well-portal $V_{0}$ [see Figs.~\ref{Fig_energyspectrum}(d) and (e)], reducing the quantum dot size $a$ leads to a shorter dephasing time ${\rm T}^{*}_{2}$ [see Figs.~\ref{Fig_T2vsa}(a) and (b)]. However, in the GaAs quantum dot, the qubit energy levels are very deep energy levels in the well, i.e, far away from the well-portal $V_{0}$ [see Fig.~\ref{Fig_energyspectrum}(f)], reducing the quantum dot size $a$ leads to a longer dephasing time ${\rm T}^{*}_{2}$ [see Fig.~\ref{Fig_T2vsa}(c)]. Note that when the qubit energy levels in the GaAs quantum dot become shallow energy levels, e.g., by tuning the quantum dot size to smaller value such as $a=30$ nm, the above discussion is no longer applicable.

\section{Summary}
In summary, in this paper we have built a theory of the spin dephasing due to the $1/f$ charge noise in a spin-orbit coupled semiconductor quantum dot. We have investigated a simple exactly solvable 1D quantum dot with both Rashba SOC and asymmetrical confining potential. Using analytical method, we obtain exactly the energy spectrum and the corresponding eigenfunctions in the quantum dot. The average values of the electric-dipole operator in the ground and the first excited states are different, such that the qubit phase noise due to the $1/f$ charge noise arises naturally in our model. It should be noted that this difference of the average values originates from the interplay between the SOC and the asymmetrical confining potential of the semiconductor quantum dot.

\section*{Acknowledgements}
This work is supported by National Natural Science Foundation of China Grant No.~11404020 and Postdoctoral Science Foundation of China Grant No.~2014M560039.

\begin{widetext}

\appendix

\section{\label{appendix_a}The transcendental equations in the strong SOC regime: $\Delta<m\alpha^{2}$}

The eigenfunction of Hamiltonian (\ref{Eq_model}) can be written as a linear combination of all the degenerate bulk wave functions, where the bulk Hamiltonian reads $H_{\rm b}=\frac{p^{2}}{2m}+\alpha\sigma^{z}p+\Delta\sigma^{x}$.
There are three types of bulk dispersion relations. The first is the plane-wave solution~\cite{LiRui2018_PRB} [see Fig.~\ref{Fig_BSStrong}(a)]
\begin{equation}
E^{\pm}_{\rm b}=\frac{k^{2}}{2m}\pm\sqrt{\alpha^{2}k^{2}+\Delta^{2}}.\label{eq_bulkspectrumI}
\end{equation}
The corresponding bulk wave functions read
\begin{equation}
\Psi^{+}_{\rm b}=\left\{\begin{array}{c}e^{ikx}\left(\begin{array}{c}\cos\frac{\theta}{2}\\\sin\frac{\theta}{2}\end{array}\right)\\
e^{-ikx}\left(\begin{array}{c}\sin\frac{\theta}{2}\\\cos\frac{\theta}{2}\end{array}\right)\end{array}\right.,
\Psi^{-}_{\rm b}=\left\{\begin{array}{c}e^{ikx}\left(\begin{array}{c}\sin\frac{\theta}{2}\\-\cos\frac{\theta}{2}\end{array}\right)\\
e^{-ikx}\left(\begin{array}{c}\cos\frac{\theta}{2}\\-\sin\frac{\theta}{2}\end{array}\right)\end{array}\right.,\label{eq_branchwave1}
\end{equation}
where $\theta\equiv\theta(k)=\arctan\left[\Delta/(\alpha\,k)\right]$. The second is the exponential-function solution~\cite{LiRui2018_PRB} [see Fig.~\ref{Fig_BSStrong}(b)]
\begin{equation}
E^{\pm}_{\rm b}=-\frac{\Gamma^{2}}{2m}\pm\sqrt{-\alpha^{2}\Gamma^{2}+\Delta^{2}}.\label{eq_bulkspectrumII}
\end{equation}
The corresponding bulk wave functions read
\begin{equation}
\Psi^{+}_{\rm b}=\left\{\begin{array}{c}e^{-\Gamma\,x}\left(\begin{array}{c}e^{i\varphi}\\1\end{array}\right)\\
e^{\Gamma\,x}\left(\begin{array}{c}e^{-i\varphi}\\1\end{array}\right)\end{array}\right.,
\Psi^{-}_{\rm b}=\left\{\begin{array}{c}e^{-\Gamma\,x}\left(\begin{array}{c}-e^{-i\varphi}\\1\end{array}\right)\\
e^{\Gamma\,x}\left(\begin{array}{c}-e^{i\varphi}\\1\end{array}\right)\end{array}\right.,\label{eq_branchwave2}
\end{equation}
where $\varphi\equiv\varphi(\Gamma)=\arctan\left(\alpha\Gamma/\sqrt{-\alpha^{2}\Gamma^{2}+\Delta^{2}}\right)$. The third is the combined plane-wave and exponential-function solution~\cite{LiRui2018_SR} [see Fig.~\ref{Fig_BSStrong}(c)]
\begin{equation}
\frac{E^{\pm}_{\rm b}}{m\alpha^{2}}=\frac{-1\pm\sqrt{(1-\frac{\Delta^{2}}{m^{2}\alpha^{4}}\sin^{2}2\phi)\cos^{2}2\phi}}{\sin^{2}2\phi}.\label{eq_bulkspectrumIII}
\end{equation}
The four degenerate bulk wave functions read
\begin{eqnarray}
\Psi^{1,3}_{\rm b}(x)&=&\left(
\begin{array}{c}
 1   \\
  R\,e^{\pm\,i\Phi}
\end{array}
\right)e^{ik_{\rho}x\cos\phi\mp\,k_{\rho}x\sin\phi},\nonumber\\
\Psi^{2,4}_{\rm b}(x)&=&\left(
\begin{array}{c}
 R\,e^{\mp\,i\Phi}   \\
  1
\end{array}
\right)e^{-ik_{\rho}x\cos\phi\mp\,k_{\rho}x\sin\phi},
\end{eqnarray}
where
\begin{eqnarray}
R\cos\Phi&=&-\frac{m\alpha^{2}+\alpha\,k_{\rho}\cos\phi}{\Delta},\nonumber\\
R\sin\Phi&=&-\frac{k^{2}_{\rho}\sin2\phi+2m\alpha\,k_{\rho}\sin\phi}{2m\Delta}.
\end{eqnarray}
In the classical allowed region $0<x<a$, the eigenfunction should be expanded using the plane-wave solutions and the exponential-function solutions, and we can divide the energy region into four sub-regions. While in the classical forbidden region $a<x$, the eigenfunction must be expanded using the combined plane-wave and exponential-function solutions.

\begin{figure}
\centering
\includegraphics{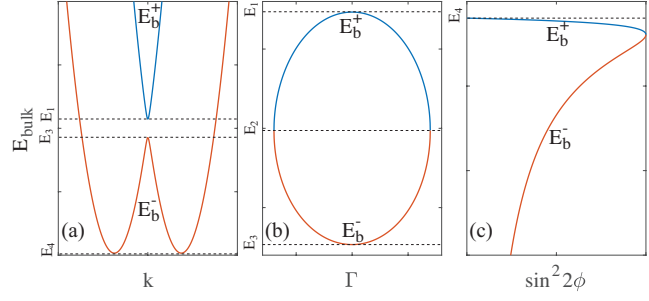}
\caption{\label{Fig_BSStrong}The bulk dispersion relations in the strong SOC regime. (a) The bulk spectrum of plane-wave solution (\ref{eq_bulkspectrumI}). (b) The bulk spectrum of exponential-function solution (\ref{eq_bulkspectrumII}). (c) The bulk spectrum of combined plane-wave and exponential-function solution (\ref{eq_bulkspectrumIII}). Here, $E_{1}=\Delta$, $E_{2}=-\frac{\Delta^{2}}{2m\alpha^{2}}$, $E_{3}=-\Delta$, and $E_{4}=-\frac{\Delta^{2}}{2m\alpha^{2}}-\frac{1}{2}m\alpha^{2}$.}
\end{figure}

\subsection{In the energy region: $-\frac{1}{2}m\alpha^{2}-\frac{\Delta^{2}}{2m\alpha^{2}}<E<-\Delta$}
In this energy region, inside the well $0<x<a$, the eigenfunction can be written as~\cite{LiRui2018_PRB}
\begin{equation}
\Psi(x)=c_{1}\left(
\begin{array}{c}
\sin\frac{\theta_{1}}{2}   \\
-\cos\frac{\theta_{1}}{2}
\end{array}
\right)e^{ik_{1}x}+c_{2}\left(
\begin{array}{c}
\cos\frac{\theta_{1}}{2}   \\
-\sin\frac{\theta_{1}}{2}
\end{array}
\right)e^{-ik_{1}x}+c_{3}\left(
\begin{array}{c}
\sin\frac{\theta_{2}}{2}   \\
-\cos\frac{\theta_{2}}{2}
\end{array}
\right)e^{ik_{2}x}+c_{4}\left(
\begin{array}{c}
\cos\frac{\theta_{2}}{2}   \\
-\sin\frac{\theta_{2}}{2}
\end{array}
\right)e^{-ik_{2}x},\label{eq_generalwfun1}
\end{equation}
where $\theta_{1,2}=\arctan[\Delta/(\alpha\,k_{1,2})]$, and $k_{1,2}$ is a function of the quantum dot energy $E$ (to be determined)
\begin{equation}
k_{1,2}=\sqrt{2}m\alpha\sqrt{1+\frac{E}{m\alpha^{2}}\mp\sqrt{1+2\frac{E}{m\alpha^{2}}+\frac{\Delta^{2}}{m^{2}\alpha^{4}}}}.
\end{equation}
Outside the well $a<x$, the eigenfunction can be written as~\cite{LiRui2018_SR}
\begin{equation}
\Psi(x)=c_{5}
\left(
\begin{array}{c}
 1   \\
R\,e^{i\Phi}
\end{array}
\right)e^{ik_{\rho}x\cos\phi-k_{\rho}x\sin\phi}+c_{6}\left(
\begin{array}{c}
 R\,e^{-i\Phi}   \\
1
\end{array}
\right)e^{-ik_{\rho}x\cos\phi-k_{\rho}x\sin\phi},\label{eq_generalwfun5}
\end{equation}
where
\begin{eqnarray}
k_{\rho}\cos\phi&=&k_{x}=m\alpha\sqrt{1+\frac{E-V_{0}}{m\alpha^{2}}+\sqrt{\frac{(E-V_{0})^{2}-\Delta^{2}}{m^{2}\alpha^{4}}}},\nonumber\\
k_{\rho}\sin\phi&=&k_{y}=m\alpha\sqrt{-1-\frac{E-V_{0}}{m\alpha^{2}}+\sqrt{\frac{(E-V_{0})^{2}-\Delta^{2}}{m^{2}\alpha^{4}}}}.\nonumber\\
\end{eqnarray}
We have six coefficients $c_{i=1,\ldots,6}$ to be determined. The boundary condition (\ref{Eq_boundary}) contains six sub-equations, such that the boundary condition can be formally written as a matrix equation $\textbf{M}\cdot\textbf{C}=0$, where the matrix $\textbf{M}$ reads
\begin{equation}
\textbf{M}=\left(
\begin{array}{cccccc}
\sin\frac{\theta_{1}}{2}  & \cos\frac{\theta_{1}}{2}  &  \sin\frac{\theta_{2}}{2} & \cos\frac{\theta_{2}}{2}& 0&0\\
\cos\frac{\theta_{1}}{2}  &  \sin\frac{\theta_{1}}{2} & \cos\frac{\theta_{2}}{2}  & \sin\frac{\theta_{2}}{2}& 0&0\\
\Lambda_{k_{1}}\sin\frac{\theta_{1}}{2}  & \Lambda_{-k_{1}}\cos\frac{\theta_{1}}{2}  & \Lambda_{k_{2}}\sin\frac{\theta_{2}}{2} & \Lambda_{-k_{2}}\cos\frac{\theta_{2}}{2}& -\Lambda_{k_{x}+ik_{y}}&-R\,e^{-i\Phi}\Lambda_{-k_{x}+ik_{y}}\\
-\Lambda_{k_{1}}\cos\frac{\theta_{1}}{2}   & -\Lambda_{-k_{1}}\sin\frac{\theta_{1}}{2}  & -\Lambda_{k_{2}}\cos\frac{\theta_{2}}{2} &-\Lambda_{-k_{2}}\sin\frac{\theta_{2}}{2} &-R~e^{i\Phi}\Lambda_{k_{x}+ik_{y}} &-\Lambda_{-k_{x}+ik_{y}}\\
 \Lambda'_{k_{1}}\sin\frac{\theta_{1}}{2}  &  \Lambda'_{-k_{1}}\cos\frac{\theta_{1}}{2} & \Lambda'_{k_{2}}\sin\frac{\theta_{2}}{2} & \Lambda'_{-k_{2}}\cos\frac{\theta_{2}}{2}&-\Lambda'_{k_{x}+ik_{y}} &-R\,e^{-i\Phi}\Lambda'_{-k_{x}+ik_{y}}\\
  -\Lambda'_{k_{1}}\cos\frac{\theta_{1}}{2}  &  -\Lambda'_{-k_{1}}\sin\frac{\theta_{1}}{2}&  -\Lambda'_{k_{2}}\cos\frac{\theta_{2}}{2}&-\Lambda'_{-k_{2}}\sin\frac{\theta_{2}}{2} & -R\,e^{i\Phi}\Lambda'_{k_{x}+ik_{y}}&-\Lambda'_{-k_{x}+ik_{y}}
\end{array}
\right).\label{eq_transc1}
\end{equation}
where $\Lambda_{k}\equiv{\rm exp}(ika)$ and $\Lambda'_{k}\equiv\,ik\times{\rm exp}(ika)$. The condition that there exists nontrivial solution for the matrix equation directly gives rise to
\begin{equation}
{\rm det}\left(\textbf{M}\right)=0.
\end{equation}
This equation is an implicit equation of the energy $E$, the solution of which gives us the energy spectrum of the quantum dot in the priorly announced energy region.

\subsection{In the energy region: $-\Delta<E<-\frac{\Delta^{2}}{2m\alpha^{2}}$}
In this energy region, inside the well $0<x<a$, the eigenfunction can be written as~\cite{LiRui2018_SR}
\begin{equation}
\Psi(x)=c_{1}e^{-\Gamma\,x}\left(\begin{array}{c}-e^{-i\varphi}\\1\end{array}\right)+c_{2}e^{\Gamma\,x}\left(\begin{array}{c}-e^{i\varphi}\\1\end{array}\right)+c_{3}e^{ikx}\left(\begin{array}{c}\sin\frac{\theta}{2}\\-\cos\frac{\theta}{2}\end{array}\right)+c_{4}e^{-ikx}\left(\begin{array}{c}\cos\frac{\theta}{2}\\-\sin\frac{\theta}{2}\end{array}\right),\label{eq_generalwfun2}
\end{equation}
where $\theta=\arctan\left[\Delta/(\alpha\,k)\right]$, $\varphi=\arctan\left(\alpha\Gamma/\sqrt{-\alpha^{2}\Gamma^{2}+\Delta^{2}}\right)$, and $k$ and $\Gamma$ are a function of the quantum dot energy $E$
\begin{eqnarray}
k&=&\sqrt{2}m\alpha\sqrt{1+\frac{E}{m\alpha^{2}}+\sqrt{1+2\frac{E}{m\alpha^{2}}+\frac{\Delta^{2}}{m^{2}\alpha^{4}}}},\nonumber\\
\Gamma&=&\sqrt{2}m\alpha\sqrt{-1-\frac{E}{m\alpha^{2}}+\sqrt{1+2\frac{E}{m\alpha^{2}}+\frac{\Delta^{2}}{m^{2}\alpha^{4}}}}.
\end{eqnarray}
Outside the well $a<x$, the eigenfunction can still be written as the form given by Eq.~(\ref{eq_generalwfun5}). In this case the matrix $\textbf{M}$ reads
\begin{equation}
\textbf{M}=\left(
\begin{array}{cccccc}
-e^{-i\varphi}  &-e^{i\varphi}  &  \sin\frac{\theta}{2} & \cos\frac{\theta}{2}& 0&0\\
1  & 1 & -\cos\frac{\theta}{2}  & -\sin\frac{\theta}{2}& 0&0\\
-e^{-\Gamma\,a-i\varphi}  & -e^{\Gamma\,a+i\varphi}  & \Lambda_{k}\sin\frac{\theta}{2} & \Lambda_{-k}\cos\frac{\theta}{2}& -\Lambda_{k_{x}+ik_{y}}&-R\,e^{-i\Phi}\Lambda_{-k_{x}+ik_{y}}\\
e^{-\Gamma\,a}   & e^{\Gamma\,a}  & -\Lambda_{k}\cos\frac{\theta}{2} &-\Lambda_{-k}\sin\frac{\theta}{2} &-R~e^{i\Phi}\Lambda_{k_{x}+ik_{y}} &-\Lambda_{-k_{x}+ik_{y}}\\
\Gamma\,e^{-\Gamma\,a-i\varphi} &  -\Gamma\,e^{\Gamma\,a+i\varphi} &\Lambda'_{k}\sin\frac{\theta}{2} & \Lambda'_{-k}\cos\frac{\theta}{2}&-\Lambda'_{k_{x}+ik_{y}} &-R\,e^{-i\Phi}\Lambda'_{-k_{x}+ik_{y}}\\
-\Gamma\,e^{-\Gamma\,a}  &   \Gamma\,e^{\Gamma\,a}&  -\Lambda'_{k}\cos\frac{\theta}{2}&-\Lambda_{-k}\sin\frac{\theta}{2} & -R\,e^{i\Phi}\Lambda'_{k_{x}+ik_{y}}&-\Lambda'_{-k_{x}+ik_{y}}
\end{array}
\right).\label{eq_transc2}
\end{equation}

\subsection{In the energy region: $-\frac{\Delta^{2}}{2m\alpha^{2}}<E<\Delta$}
In this energy region, inside the well $0<x<a$, the eigenfunction can be written as~\cite{LiRui2018_PRB}
\begin{equation}
\Psi(x)=c_{1}e^{-\Gamma\,x}\left(\begin{array}{c}e^{i\varphi}\\1\end{array}\right)+c_{2}e^{\Gamma\,x}\left(\begin{array}{c}e^{-i\varphi}\\1\end{array}\right)+c_{3}e^{ikx}\left(\begin{array}{c}\sin\frac{\theta}{2}\\-\cos\frac{\theta}{2}\end{array}\right)+c_{4}e^{-ikx}\left(\begin{array}{c}\cos\frac{\theta}{2}\\-\sin\frac{\theta}{2}\end{array}\right).\label{eq_generalwfun3}
\end{equation}
Outside the well $a<x$, the eigenfunction can still be written as the form given by Eq.~(\ref{eq_generalwfun5}). In this case the matrix $\textbf{M}$ reads
\begin{equation}
\textbf{M}=\left(
\begin{array}{cccccc}
e^{i\varphi}  &e^{-i\varphi}  &  \sin\frac{\theta}{2} & \cos\frac{\theta}{2}& 0&0\\
1  & 1 & -\cos\frac{\theta}{2}  & -\sin\frac{\theta}{2}& 0&0\\
e^{-\Gamma\,a+i\varphi}  & e^{\Gamma\,a-i\varphi}  & \Lambda_{k}\sin\frac{\theta}{2} & \Lambda_{-k}\cos\frac{\theta}{2}& -\Lambda_{k_{x}+ik_{y}}&-R\,e^{-i\Phi}\Lambda_{-k_{x}+ik_{y}}\\
e^{-\Gamma\,a}   & e^{\Gamma\,a}  & -\Lambda_{k}\cos\frac{\theta}{2} &-\Lambda_{-k}\sin\frac{\theta}{2} &-R~e^{i\Phi}\Lambda_{k_{x}+ik_{y}} &-\Lambda_{-k_{x}+ik_{y}}\\
-\Gamma\,e^{-\Gamma\,a+i\varphi} &  \Gamma\,e^{\Gamma\,a-i\varphi} & \Lambda'_{k}\sin\frac{\theta}{2} & \Lambda'_{-k}\cos\frac{\theta}{2}&-\Lambda'_{k_{x}+ik_{y}} &-R\,e^{-i\Phi}\Lambda'_{-k_{x}+ik_{y}}\\
-\Gamma\,e^{-\Gamma\,a}  &   \Gamma\,e^{\Gamma\,a}&  -\Lambda'_{k}\cos\frac{\theta}{2}&-\Lambda'_{-k}\sin\frac{\theta}{2} & -R\,e^{i\Phi}\Lambda'_{k_{x}+ik_{y}}&-\Lambda'_{-k_{x}+ik_{y}}
\end{array}
\right).\label{eq_transc3}
\end{equation}

\subsection{In the energy region: $\Delta<E$}
In this energy region, inside the well $0<x<a$, the eigenfunction can be written as~\cite{LiRui2018_PRB}
 \begin{equation}
\Psi(x)=c_{1}\left(
\begin{array}{c}
\cos\frac{\theta_{1}}{2}   \\
\sin\frac{\theta_{1}}{2}
\end{array}
\right)e^{ik_{1}x}+c_{2}\left(
\begin{array}{c}
\sin\frac{\theta_{1}}{2}   \\
\cos\frac{\theta_{1}}{2}
\end{array}
\right)e^{-ik_{1}x}+c_{3}\left(
\begin{array}{c}
\sin\frac{\theta_{2}}{2}   \\
-\cos\frac{\theta_{2}}{2}
\end{array}
\right)e^{ik_{2}x}+c_{4}\left(
\begin{array}{c}
\cos\frac{\theta_{2}}{2}   \\
-\sin\frac{\theta_{2}}{2}
\end{array}
\right)e^{-ik_{2}x}.\label{eq_generalwfun4}
\end{equation}
Outside the well $a<x$, the eigenfunction can still be written as the form given by Eq.~(\ref{eq_generalwfun5}). In this case the matrix $\textbf{M}$ reads
\begin{equation}
\textbf{M}=\left(
\begin{array}{cccccc}
\cos\frac{\theta_{1}}{2}  & \sin\frac{\theta_{1}}{2}  &  \sin\frac{\theta_{2}}{2} & \cos\frac{\theta_{2}}{2}& 0&0\\
\sin\frac{\theta_{1}}{2}  &  \cos\frac{\theta_{1}}{2} & -\cos\frac{\theta_{2}}{2}  & -\sin\frac{\theta_{2}}{2}& 0&0\\
\Lambda_{k_{1}}\cos\frac{\theta_{1}}{2}  & \Lambda_{-k_{1}}\sin\frac{\theta_{1}}{2}  & \Lambda_{k_{2}}\sin\frac{\theta_{2}}{2} & \Lambda_{-k_{2}}\cos\frac{\theta_{2}}{2}& -\Lambda_{k_{x}+ik_{y}}&-R~e^{-i\Phi}\Lambda_{-k_{x}+ik_{y}}\\
\Lambda_{k_{1}}\sin\frac{\theta_{1}}{2}   & \Lambda_{-k_{1}}\cos\frac{\theta_{1}}{2}  & -\Lambda_{k_{2}}\cos\frac{\theta_{2}}{2} &-\Lambda_{-k_{2}}\sin\frac{\theta_{2}}{2} &-R~e^{i\Phi}\Lambda_{k_{x}+ik_{y}} &-\Lambda_{-k_{x}+ik_{y}}\\
\Lambda'_{k_{1}}\cos\frac{\theta_{1}}{2}  &  \Lambda'_{-k_{1}}\sin\frac{\theta_{1}}{2} & \Lambda'_{k_{2}}\sin\frac{\theta_{2}}{2} & \Lambda'_{-k_{2}}\cos\frac{\theta_{2}}{2}&-\Lambda'_{k_{x}+ik_{y}} &-R~e^{-i\Phi}\Lambda'_{-k_{x}+ik_{y}}\\
\Lambda'_{k_{1}}\sin\frac{\theta_{1}}{2}  &   \Lambda'_{-k_{1}}\cos\frac{\theta_{1}}{2}&  -\Lambda'_{k_{2}}\cos\frac{\theta_{2}}{2}&-\Lambda'_{-k_{2}}\sin\frac{\theta_{2}}{2} & -R~e^{i\Phi}\Lambda'_{k_{x}+ik_{y}}&-\Lambda'_{-k_{x}+ik_{y}}
\end{array}
\right).\label{eq_transc4}
\end{equation}

\section{\label{appendix_b}The transcendental equations in the intermediate SOC regime: $\frac{\Delta}{2}<m\alpha^{2}<\Delta$}

\begin{figure}
\centering
\includegraphics{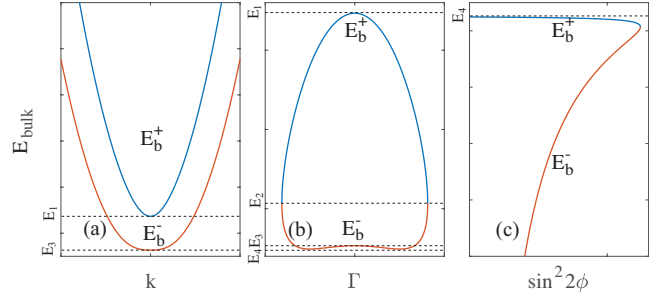}
\caption{\label{Fig_BSInterme}The bulk dispersion relations in the intermediate SOC regime. (a) The bulk spectrum of plane-wave solution (\ref{eq_bulkspectrumI}). (b) The bulk spectrum of exponential-function solution (\ref{eq_bulkspectrumII}). (c) The bulk spectrum of combined plane-wave and exponential-function solution (\ref{eq_bulkspectrumIII}). Here, $E_{1}=\Delta$, $E_{2}=-\frac{\Delta^{2}}{2m\alpha^{2}}$, $E_{3}=-\Delta$, and $E_{4}=-\frac{\Delta^{2}}{2m\alpha^{2}}-\frac{1}{2}m\alpha^{2}$.}
\end{figure}

In the intermediate SOC regime, the bulk dispersion relations are shown in Fig.~\ref{Fig_BSInterme}. In the classical allowed region $x<a$, the energy region is divide into three sub-regions. While in the classical forbidden region $a<x$, the energy region is divided into two sub-regions.

\subsection{In the energy region: $-\Delta<E<-\frac{\Delta^{2}}{2m\alpha^{2}}$~$\bigcap$~$E-V_{0}<-\frac{\Delta^{2}}{2m\alpha^{2}}-\frac{1}{2}m\alpha^{2}$}
In this case the Matrix $\textbf{M}$ is the same as that given by Eq.~(\ref{eq_transc2}).

\subsection{In the energy region: $-
\Delta<E<-\frac{\Delta^{2}}{2m\alpha^{2}}$~$\bigcap$~$-\frac{\Delta^{2}}{2m\alpha^{2}}-\frac{1}{2}m\alpha^{2}<E-V_{0}<-
\Delta$}
In this energy region, inside the well $0<x<a$, the eigenfunction is still written as the form given by  Eq.~(\ref{eq_generalwfun2}). Outside the well $a<x$, the eigenfunction reads
\begin{equation}
\Psi(x)=c_{5}e^{-\Gamma_{1}x}
\left(
\begin{array}{c}
 -e^{-i\varphi_{1}}   \\
  1
\end{array}
\right)+c_{6}e^{-\Gamma_{2}x}\left(
\begin{array}{c}
 -e^{-i\varphi_{2}}   \\
  1
\end{array}
\right),\label{eq_generalwfun6}
\end{equation}
where $\varphi_{1,2}=\arctan(\alpha\Gamma_{1,2}/\sqrt{-\alpha^{2}\Gamma^{2}_{1,2}+\Delta^{2}})$, and $\Gamma_{1,2}$ are a function of the quantum dot energy $E$ to be determined
\begin{equation}
\Gamma_{1,2}=\sqrt{2}m\alpha\sqrt{-1-\frac{E-V_{0}}{m\alpha^{2}}\pm\sqrt{1+\frac{2(E-V_{0})}{m\alpha^{2}}+\frac{\Delta^{2}}{m^{2}\alpha^{4}}}}.
\end{equation}
In this case the matrix $\textbf{M}$ reads
\begin{equation}
\textbf{M}=\left(
\begin{array}{cccccc}
-e^{-i\varphi}  &-e^{i\varphi}  &  \sin\frac{\theta}{2} & \cos\frac{\theta}{2}& 0&0\\
1  & 1 & -\cos\frac{\theta}{2}  & -\sin\frac{\theta}{2}& 0&0\\
-e^{-\Gamma\,a-i\varphi}  & -e^{\Gamma\,a+i\varphi}  & \Lambda_{k}\sin\frac{\theta}{2} & \Lambda_{-k}\cos\frac{\theta}{2}& e^{-\Gamma_{1}a-i\varphi_{1}}&e^{-\Gamma_{2}a-i\varphi_{2}}\\
e^{-\Gamma\,a}   & e^{\Gamma\,a}  & -\Lambda_{k}\cos\frac{\theta}{2} &-\Lambda_{-k}\sin\frac{\theta}{2} &-e^{-\Gamma_{1}a} &-e^{-\Gamma_{2}a}\\
\Gamma\,e^{-\Gamma\,a-i\varphi} &  -\Gamma\,e^{\Gamma\,a+i\varphi} & \Lambda'_{k}\sin\frac{\theta}{2} & \Lambda'_{-k}\cos\frac{\theta}{2}&-\Gamma_{1}e^{-\Gamma_{1}a-i\varphi_{1}} &-\Gamma_{2}e^{-\Gamma_{2}a-i\varphi_{2}}\\
-\Gamma\,e^{-\Gamma\,a}  &   \Gamma\,e^{\Gamma\,a}&  -\Lambda'_{k}\cos\frac{\theta}{2}&-\Lambda'_{-k}\sin\frac{\theta}{2} & \Gamma_{1}e^{-\Gamma_{1}a}&\Gamma_{2}e^{-\Gamma_{2}a}\\
\end{array}
\right).\label{eq_transc5}
\end{equation}

\subsection{In the energy region: $-\frac{\Delta^{2}}{2m\alpha^{2}}<E<
\Delta$~$\bigcap$~$E-V_{0}<-\frac{\Delta^{2}}{2m\alpha^{2}}-\frac{1}{2}m\alpha^{2}$}
In this case the Matrix $\textbf{M}$ is the same as that given by Eq.~(\ref{eq_transc3}).

\subsection{In the energy region: $-\frac{\Delta^{2}}{2m\alpha^{2}}<E<
\Delta$~$\bigcap$~$-\frac{\Delta^{2}}{2m\alpha^{2}}-\frac{1}{2}m\alpha^{2}<E-V_{0}<-
\Delta$}
In this energy region, inside the well $0<x<a$, the eigenfunction is written as the form given by Eq.~(\ref{eq_generalwfun3}). Outside the well $a<x$, the eigenfunction is written as the form given by Eq.~(\ref{eq_generalwfun6}). In this case the matrix $\textbf{M}$ reads
\begin{equation}
\textbf{M}=\left(
\begin{array}{cccccc}
e^{i\varphi}  &e^{-i\varphi}  &  \sin\frac{\theta}{2} & \cos\frac{\theta}{2}& 0&0\\
1  & 1 & -\cos\frac{\theta}{2}  & -\sin\frac{\theta}{2}& 0&0\\
e^{-\Gamma\,a+i\varphi}  & e^{\Gamma\,a-i\varphi}  & \Lambda_{k}\sin\frac{\theta}{2} & \Lambda_{-k}\cos\frac{\theta}{2}& e^{-\Gamma_{1}a-i\varphi_{1}}&e^{-\Gamma_{2}a-i\varphi_{2}}\\
e^{-\Gamma\,a}   & e^{\Gamma\,a}  & -\Lambda_{k}\cos\frac{\theta}{2} &-\Lambda_{-k}\sin\frac{\theta}{2} &-e^{-\Gamma_{1}a} &-e^{-\Gamma_{2}a}\\
-\Gamma\,e^{-\Gamma\,a+i\varphi} &  \Gamma\,e^{\Gamma\,a-i\varphi} & \Lambda'_{k}\sin\frac{\theta}{2} & \Lambda'_{-k}\cos\frac{\theta}{2}&-\Gamma_{1}e^{-\Gamma_{1}a-i\varphi_{1}} &-\Gamma_{2}e^{-\Gamma_{2}a-i\varphi_{2}}\\
-\Gamma\,e^{-\Gamma\,a}  &   \Gamma\,e^{\Gamma\,a}&  -\Lambda'_{k}\cos\frac{\theta}{2}&-\Lambda'_{-k}\sin\frac{\theta}{2} & \Gamma_{1}e^{-\Gamma_{1}a}&\Gamma_{2}e^{-\Gamma_{2}a}\\
\end{array}
\right).\label{eq_transc6}
\end{equation}

\subsection{In the energy region: $\Delta<E$~$\bigcap$~$E-V_{0}<-\frac{\Delta^{2}}{2m\alpha^{2}}-\frac{1}{2}m\alpha^{2}$}
In this case the Matrix $\textbf{M}$ is the same as that given by Eq.~(\ref{eq_transc4}).

\subsection{In the energy region: $\Delta<E$~$\bigcap$~$-\frac{\Delta^{2}}{2m\alpha^{2}}-\frac{1}{2}m\alpha^{2}<E-V_{0}<-
\Delta$}
In this energy region, inside the well $0<x<a$, the eigenfunction is written as the form given by Eq.~(\ref{eq_generalwfun4}). Outside the well $a<x$, the eigenfunction is written as the form given by Eq.~(\ref{eq_generalwfun6}). In this case the matrix $\textbf{M}$ reads
\begin{equation}
\textbf{M}=\left(
\begin{array}{cccccc}
\cos\frac{\theta_{1}}{2}  & \sin\frac{\theta_{1}}{2}  &  \sin\frac{\theta_{2}}{2} & \cos\frac{\theta_{2}}{2}& 0&0\\
\sin\frac{\theta_{1}}{2}  &  \cos\frac{\theta_{1}}{2} & -\cos\frac{\theta_{2}}{2}  & -\sin\frac{\theta_{2}}{2}& 0&0\\
\Lambda_{k_{1}}\cos\frac{\theta_{1}}{2}  & \Lambda_{-k_{1}}\sin\frac{\theta_{1}}{2}  & \Lambda_{k_{2}}\sin\frac{\theta_{2}}{2} & \Lambda_{-k_{2}}\cos\frac{\theta_{2}}{2}&e^{-\Gamma_{1}a-i\varphi_{1}}&e^{-\Gamma_{2}a-i\varphi_{2}}\\
\Lambda_{k_{1}}\sin\frac{\theta_{1}}{2}   & \Lambda_{-k_{1}}\cos\frac{\theta_{1}}{2}  & -\Lambda_{k_{2}}\cos\frac{\theta_{2}}{2} &-\Lambda_{-k_{2}}\sin\frac{\theta_{2}}{2} &-e^{-\Gamma_{1}a} &-e^{-\Gamma_{2}a}\\
\Lambda'_{k_{1}}\cos\frac{\theta_{1}}{2}  &  \Lambda'_{-k_{1}}\sin\frac{\theta_{1}}{2} & \Lambda'_{k_{2}}\sin\frac{\theta_{2}}{2} & \Lambda'_{-k_{2}}\cos\frac{\theta_{2}}{2}&-\Gamma_{1}e^{-\Gamma_{1}a-i\varphi_{1}} &-\Gamma_{2}e^{-\Gamma_{2}a-i\varphi_{2}}\\
\Lambda'_{k_{1}}\sin\frac{\theta_{1}}{2}  &  \Lambda'_{-k_{1}}\cos\frac{\theta_{1}}{2}&  -\Lambda'_{k_{2}}\cos\frac{\theta_{2}}{2}&-\Lambda'_{-k_{2}}\sin\frac{\theta_{2}}{2} &\Gamma_{1}e^{-\Gamma_{1}a}&\Gamma_{2}e^{-\Gamma_{2}a}
\end{array}
\right).\label{eq_transc7}
\end{equation}

\section{\label{appendix_c}The transcendental equations in the weak SOC regime: $m\alpha^{2}<\frac{\Delta}{2}$}

\begin{figure}
\centering
\includegraphics{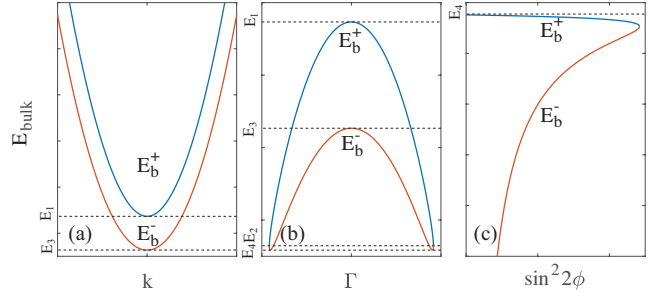}
\caption{\label{Fig_BSWeak}The bulk dispersion relations in the weak SOC regime. (a) The bulk spectrum of plane-wave solution (\ref{eq_bulkspectrumI}). (b) The bulk spectrum of exponential-function solution (\ref{eq_bulkspectrumII}). (c) The bulk spectrum of combined plane-wave and exponential-function solution (\ref{eq_bulkspectrumIII}). Here, $E_{1}=\Delta$, $E_{2}=-\frac{\Delta^{2}}{2m\alpha^{2}}$, $E_{3}=-\Delta$, and $E_{4}=-\frac{\Delta^{2}}{2m\alpha^{2}}-\frac{1}{2}m\alpha^{2}$.}
\end{figure}

In the weak SOC regime, the bulk dispersion relations are shown in Fig.~\ref{Fig_BSWeak}. In the classical allowed region $x<a$, the energy region is divide into two sub-regions. While in the classical forbidden region $a<x$, the energy region is divided into three sub-regions.

\subsection{In the energy region: $-\Delta<E<\Delta$~$\bigcap$~$E-V_{0}<-\frac{\Delta^{2}}{2m\alpha^{2}}-\frac{1}{2}m\alpha^{2}$}
In this case the Matrix $\textbf{M}$ is the same as that given by Eq.~(\ref{eq_transc3}).

\subsection{In the energy region: $-\Delta<E<\Delta$~$\bigcap$~$-\frac{\Delta^{2}}{2m\alpha^{2}}-\frac{1}{2}m\alpha^{2}<E-V_{0}<-\frac{\Delta^{2}}{2m\alpha^{2}}$}
In this case the Matrix $\textbf{M}$ is the same as that given by Eq.~(\ref{eq_transc6}).

\subsection{In the energy region: $-\Delta<E<\Delta$~$\bigcap$~$-\frac{\Delta^{2}}{2m\alpha^{2}}<E-V_{0}<-\Delta$}
In this energy region, inside the well $0<x<a$, the eigenfunction is written as the form given by Eq.~(\ref{eq_generalwfun3}). Outside the well $a<x$, the eigenfunction reads
\begin{equation}
\Psi(x)=c_{5}e^{-\Gamma_{1}x}
\left(
\begin{array}{c}
 e^{i\varphi_{1}}   \\
  1
\end{array}
\right)+c_{6}e^{-\Gamma_{2}x}\left(
\begin{array}{c}
 -e^{-i\varphi_{2}}   \\
  1
\end{array}
\right),\label{eq_generalwfun7}
\end{equation}
In this case the matrix $\textbf{M}$ reads
\begin{equation}
\textbf{M}=\left(
\begin{array}{cccccc}
e^{i\varphi}  &e^{-i\varphi}  &  \sin\frac{\theta}{2} & \cos\frac{\theta}{2}& 0&0\\
1  & 1 & -\cos\frac{\theta}{2}  & -\sin\frac{\theta}{2}& 0&0\\
e^{-\Gamma\,a+i\varphi}  & e^{\Gamma\,a-i\varphi}  & \Lambda_{k}\sin\frac{\theta}{2} & \Lambda_{-k}\cos\frac{\theta}{2}& -e^{-\Gamma_{1}a+i\varphi_{1}}&e^{-\Gamma_{2}a-i\varphi_{2}}\\
e^{-\Gamma\,a}   & e^{\Gamma\,a}  & -\Lambda_{k}\cos\frac{\theta}{2} &-\Lambda_{-k}\sin\frac{\theta}{2} &-e^{-\Gamma_{1}a} &-e^{-\Gamma_{2}a}\\
-\Gamma\,e^{-\Gamma\,a+i\varphi} &  \Gamma\,e^{\Gamma\,a-i\varphi} & \Lambda'_{k}\sin\frac{\theta}{2} & \Lambda'_{-k}\cos\frac{\theta}{2}&\Gamma_{1}e^{-\Gamma_{1}a+i\varphi_{1}} &-\Gamma_{2}e^{-\Gamma_{2}a-i\varphi_{2}}\\
-\Gamma\,e^{-\Gamma\,a}  &   \Gamma\,e^{\Gamma\,a}&  -\Lambda'_{k}\cos\frac{\theta}{2}&-\Lambda'_{-k}\sin\frac{\theta}{2} & \Gamma_{1}e^{-\Gamma_{1}a}&\Gamma_{2}e^{-\Gamma_{2}a}\\
\end{array}
\right).\label{eq_transc8}
\end{equation}

\subsection{In the energy region: $\Delta<E$~$\bigcap$~$E-V_{0}<-\frac{\Delta^{2}}{2m\alpha^{2}}-\frac{1}{2}m\alpha^{2}$}
In this case the Matrix $\textbf{M}$ is the same as that given by Eq.~(\ref{eq_transc4}).

\subsection{In the energy region: $\Delta<E$~$\bigcap$~$-\frac{\Delta^{2}}{2m\alpha^{2}}-\frac{1}{2}m\alpha^{2}<E-V_{0}<-\frac{\Delta^{2}}{2m\alpha^{2}}$}
In this case the Matrix $\textbf{M}$ is the same as that given by Eq.~(\ref{eq_transc7}).

\subsection{In the energy region: $\Delta<E$~$\bigcap$~$-\frac{\Delta^{2}}{2m\alpha^{2}}<E-V_{0}<-\Delta$}
In this energy region, inside the well $0<x<a$, the eigenfunction is written as the form given by Eq.~(\ref{eq_generalwfun4}). Outside the well $a<x$, the eigenfunction is written as the form given by Eq.~(\ref{eq_generalwfun7}). In this case the matrix $\textbf{M}$ reads
\begin{equation}
\textbf{M}=\left(
\begin{array}{cccccc}
\cos\frac{\theta_{1}}{2}  & \sin\frac{\theta_{1}}{2}  &  \sin\frac{\theta_{2}}{2} & \cos\frac{\theta_{2}}{2}& 0&0\\
\sin\frac{\theta_{1}}{2}  &  \cos\frac{\theta_{1}}{2} & -\cos\frac{\theta_{2}}{2}  & -\sin\frac{\theta_{2}}{2}& 0&0\\
\Lambda_{k_{1}}\cos\frac{\theta_{1}}{2}  & \Lambda_{-k_{1}}\sin\frac{\theta_{1}}{2}  & \Lambda_{k_{2}}\sin\frac{\theta_{2}}{2} & \Lambda_{-k_{2}}\cos\frac{\theta_{2}}{2}&-e^{-\Gamma_{1}a+i\varphi_{1}}&e^{-\Gamma_{2}a-i\varphi_{2}}\\
\Lambda_{k_{1}}\sin\frac{\theta_{1}}{2}   & \Lambda_{-k_{1}}\cos\frac{\theta_{1}}{2}  & -\Lambda_{k_{2}}\cos\frac{\theta_{2}}{2} &-\Lambda_{-k_{2}}\sin\frac{\theta_{2}}{2} &-e^{-\Gamma_{1}a} &-e^{-\Gamma_{2}a}\\
\Lambda'_{k_{1}}\cos\frac{\theta_{1}}{2}  &  \Lambda'_{-k_{1}}\sin\frac{\theta_{1}}{2} & \Lambda'_{k_{2}}\sin\frac{\theta_{2}}{2} & \Lambda'_{-k_{2}}\cos\frac{\theta_{2}}{2}&\Gamma_{1}e^{-\Gamma_{1}a+i\varphi_{1}} &-\Gamma_{2}e^{-\Gamma_{2}a-i\varphi_{2}}\\
\Lambda'_{k_{1}}\sin\frac{\theta_{1}}{2}  &   \Lambda'_{-k_{1}}\cos\frac{\theta_{1}}{2}&  -\Lambda'_{k_{2}}\cos\frac{\theta_{2}}{2}&-\Lambda'_{-k_{2}}\sin\frac{\theta_{2}}{2} &\Gamma_{1}e^{-\Gamma_{1}a}&\Gamma_{2}e^{-\Gamma_{2}a}
\end{array}
\right).\label{eq_transc9}
\end{equation}

\section{\label{appendix_d}The derivation of the spin dephasing rate}
For the derivation convenience, here we write the Hamiltonian of the qubit-environment (noise) interaction again
\begin{equation}
H_{\rm tot}=\frac{E_{\rm e}-E_{\rm g}}{2}\tau^{z}+\sum_{k}\hbar\omega_{k}b^{\dagger}_{k}b_{k}+\sum_{k}\left(\frac{\chi^{\rm e}_{k}+\chi^{\rm g}_{k}}{2}+\frac{\chi^{\rm e}_{k}-\chi^{\rm g}_{k}}{2}\tau^{z}\right)(b_{k}+b^{\dagger}_{k}),
\end{equation}
where $\chi^{\rm e/g}_{k}=ex_{\rm e/g}\Xi_{k}\cos\Theta$. For the time evolution problem, the reduced density matrix of the spin qubit can always written as
\begin{equation}
\rho_{\rm q}(t)={\rm Tr}_{\rm en}\{\rho_{\rm tot}(t)\}=\rho_{\Uparrow\Uparrow}(t)|\!\!\Uparrow\rangle\langle\Uparrow\!\!|+\rho_{\Uparrow\Downarrow}(t)|\!\!\Uparrow\rangle\langle\Downarrow\!\!|+\rho_{\Downarrow\Uparrow}(t)|\!\!\Downarrow\rangle\langle\Uparrow\!\!|+\rho_{\Downarrow\Downarrow}(t)|\!\!\Downarrow\rangle\langle\Downarrow\!\!|,
\end{equation}
where $\rho_{\rm tot}(t)$ is the density matrix of the total qubit-environment system. From the above equation, the off-diagonal element of the reduced density matrix, which is used to quantify the qubit phase coherence, can be written as
\begin{equation}
\rho_{\Uparrow\Downarrow}(t)={\rm Tr}_{\rm q}\left\{|\!\!\Downarrow\rangle\langle\Uparrow\!\!|\rho_{\rm q}(t)\right\}={\rm Tr}_{\rm tot}\left\{|\!\!\Downarrow\rangle\langle\Uparrow\!\!|\rho_{\rm tot}(t)\right\}.
\end{equation}
Initially, the qubit-environment system is in a product state
\begin{equation}
\rho_{\rm tot}(0)=\rho_{\rm q}(0)\otimes\rho_{\rm en}(0),
\end{equation}
where the environment is in a thermal state $\rho_{\rm en}(0)=\prod_{k}\rho^{k}_{\rm en}=\prod_{k}(1-e^{-\beta\hbar\omega_{k}})e^{-\beta\hbar\omega_{k}b^{\dagger}_{k}b_{k}}$, with $\beta=1/(k_{B}T)$. It is convenient to consider the qubit dephasing in the Heisenberg picture
\begin{eqnarray}
\rho_{\Uparrow\Downarrow}(t)&=&{\rm Tr}_{\rm tot}\left\{e^{iH_{\rm tot}t/\hbar}|\!\!\Downarrow\rangle\langle\Uparrow\!\!|e^{-iH_{\rm tot}t/\hbar}\rho_{\rm tot}(0)\right\}\nonumber\\
&=&{\rm Tr}_{\rm tot}\left\{e^{iH_{\rm tot}t/\hbar}|\!\!\Downarrow\rangle\langle\Uparrow\!\!|e^{-iH_{\rm tot}t/\hbar}\rho_{\Uparrow\Downarrow}(0)|\!\!\Uparrow\rangle\langle\Downarrow\!\!|\otimes\rho_{\rm en}(0)\right\}\nonumber\\
&=&\rho_{\Uparrow\Downarrow}(0){\rm Tr}_{\rm en}\left\{e^{iH_{\Downarrow}t/\hbar}e^{-iH_{\Uparrow}t/\hbar}\rho_{\rm en}(0)\right\},\label{eq_phasecoherence}
\end{eqnarray}
where
\begin{eqnarray}
H_{\Uparrow/\Downarrow}&=&\pm\frac{E_{\rm e}-E_{\rm g}}{2}+\sum_{k}\hbar\omega_{k}b^{\dagger}_{k}b_{k}+\sum_{k}\chi^{\rm e/g}_{k}(b_{k}+b^{\dagger}_{k})\nonumber\\
&=&\pm\frac{E_{\rm e}-E_{\rm g}}{2}+\sum_{k}\hbar\omega_{k}\left(b^{\dagger}_{k}+\frac{\chi^{\rm e/g}_{k}}{\hbar\omega_{k}}\right)\left(b_{k}+\frac{\chi^{\rm e/g}_{k}}{\hbar\omega_{k}}\right)-\sum_{k}\frac{|\chi^{\rm e/g}_{k}|^{2}}{\hbar\omega_{k}}.
\end{eqnarray}
Substituting $H_{\Uparrow/\Downarrow}$ in the last line of Eq.~(\ref{eq_phasecoherence}) with the above expressions, we obtain
\begin{eqnarray}
\rho_{\Uparrow\Downarrow}(t)
&=&\rho_{\Uparrow\Downarrow}(0)\prod_{k}{\rm Tr}\left\{{\rm exp}\left[i\omega_{k}t\left(b^{\dagger}_{k}+\frac{\chi^{\rm g}_{k}}{\hbar\omega_{k}}\right)\left(b_{k}+\frac{\chi^{\rm g}_{k}}{\hbar\omega_{k}}\right)\right]{\rm exp}\left[-i\omega_{k}t\left(b^{\dagger}_{k}+\frac{\chi^{\rm e}_{k}}{\hbar\omega_{k}}\right)\left(b_{k}+\frac{\chi^{\rm e}_{k}}{\hbar\omega_{k}}\right)\right]\rho^{k}_{\rm en}\right\}.
\end{eqnarray}
Introducing the displacement operator $D_{k}(\beta)={\rm exp}\left(\beta\,b^{\dagger}_{k}-\beta^{*}b_{k}\right)$, we have
\begin{eqnarray}
&&D^{\dagger}_{k}\left(\frac{\chi^{\rm g}_{k}}{\hbar\omega_{k}}\right)e^{i\omega_{k}tb^{\dagger}_{k}b_{k}}D_{k}\left(\frac{\chi^{\rm g}_{k}}{\hbar\omega_{k}}\right)={\rm exp}\left[i\omega_{k}t\left(b^{\dagger}_{k}+\frac{\chi^{\rm g}_{k}}{\hbar\omega_{k}}\right)\left(b_{k}+\frac{\chi^{\rm g}_{k}}{\hbar\omega_{k}}\right)\right],\nonumber\\
&&D^{\dagger}_{k}\left(\frac{\chi^{\rm e}_{k}}{\hbar\omega_{k}}\right)e^{-i\omega_{k}tb^{\dagger}_{k}b_{k}}D_{k}\left(\frac{\chi^{\rm e}_{k}}{\hbar\omega_{k}}\right)={\rm exp}\left[-i\omega_{k}t\left(b^{\dagger}_{k}+\frac{\chi^{\rm e}_{k}}{\hbar\omega_{k}}\right)\left(b_{k}+\frac{\chi^{\rm e}_{k}}{\hbar\omega_{k}}\right)\right].
\end{eqnarray}
Therefore, the off-diagonal element $\rho_{\Uparrow\Downarrow}(t)$ reads
\begin{eqnarray}
\rho_{\Uparrow\Downarrow}(t)&=&\rho_{\Uparrow\Downarrow}(0)\prod_{k}{\rm Tr}\left\{D^{\dagger}_{k}\left(\frac{\chi^{\rm g}_{k}}{\hbar\omega_{k}}\right)e^{i\omega_{k}tb^{\dagger}_{k}b_{k}}D_{k}\left(-\frac{\chi^{\rm e}_{k}-\chi^{\rm g}_{k}}{\hbar\omega_{k}}\right)e^{-i\omega_{k}tb^{\dagger}_{k}b_{k}}D_{k}\left(\frac{\chi^{\rm e}_{k}}{\hbar\omega_{k}}\right)\rho^{k}_{\rm en}\right\}\nonumber\\
&=&\rho_{\Uparrow\Downarrow}(0)\prod_{k}{\rm Tr}\left\{D^{\dagger}_{k}\left(\frac{\chi^{\rm g}_{k}}{\hbar\omega_{k}}\right)D_{k}\left(-\frac{\chi^{\rm e}_{k}-\chi^{\rm g}_{k}}{\hbar\omega_{k}}e^{i\omega_{k}t}\right)D_{k}\left(\frac{\chi^{\rm e}_{k}}{\hbar\omega_{k}}\right)\rho^{k}_{\rm en}\right\}\nonumber\\
&=&\rho_{\Uparrow\Downarrow}(0)\prod_{k}{\rm Tr}\left\{D_{k}\left[\frac{(\chi^{\rm e}_{k}-\chi^{\rm g}_{k})}{\hbar\omega_{k}}(1-e^{i\omega_{k}t})\right]\rho^{k}_{\rm en}\right\}\nonumber\\
&=&\rho_{\Uparrow\Downarrow}(0)\prod_{k}{\rm exp}\left\{-2\frac{(x_{\rm e}-x_{\rm g})^{2}e^{2}\Xi^{2}_{k}\cos^{2}\Theta\sin^{2}\frac{\omega_{k}t}{2}}{\hbar^{2}\omega^{2}_{k}}[2n(\omega_{k})+1]\right\}.
\end{eqnarray}
In deriving the last line of the above equation, we have used the property ${\rm Tr}\left\{D_{k}(\beta)\rho^{k}_{\rm en}(0)\right\}={\rm exp}(-|\beta|^{2}/2){\rm Tr}\left\{{\rm exp}(\beta\,b^{\dagger}_{k}){\rm exp}(-\beta^{*}b_{k})\rho^{k}_{\rm en}(0)\right\}={\rm exp}\left\{-|\beta|^{2}\left[n(\omega_{k})+1/2\right]\right\}$~\cite{Gardiner1991}.
If we model the qubit phase coherence as $\left|\rho_{\Uparrow\Downarrow}(t)/\rho_{\Uparrow\Downarrow}(0)\right|={\rm exp}[-\Gamma_{\rm ph}(t)]$, the dephasing rate can be written as
\begin{eqnarray}
\Gamma_{\rm ph}(t)&\equiv&\langle\Gamma_{\rm ph}(t)\rangle_{\Theta}=\sum_{k}\frac{(x_{\rm e}-x_{\rm g})^{2}e^{2}\Xi^{2}_{k}\sin^{2}\frac{\omega_{k}t}{2}}{\hbar^{2}\omega^{2}_{k}}[2n(\omega_{k})+1]\nonumber\\
&=&\sum_{k}\int\,d\omega\frac{(x_{\rm e}-x_{\rm g})^{2}e^{2}\Xi^{2}(\omega)\sin^{2}\frac{\omega\,t}{2}}{\hbar^{2}\omega^{2}}[2n(\omega)+1]\delta(\omega-\omega_{k})\nonumber\\
&=&\frac{(x_{\rm e}-x_{\rm g})^{2}}{2a^{2}}\int\,d\omega\,S(\omega)\frac{\sin^{2}(\omega\,t/2)}{(\omega/2)^{2}},
\end{eqnarray}
where the spectrum function is defined as
\begin{equation}
S(\omega)=\sum_{k}\frac{e^{2}a^{2}\Xi^{2}(\omega)[2n(\omega)+1]}{2\hbar^{2}}\delta(\omega-\omega_{k}).
\end{equation}
We are only interested in the low-frequency $1/f$ noise, where $n(\omega)=1/\left[{\rm exp}(\hbar\omega/k_{B}T)-1\right]\approx\,k_{B}T/(\hbar\omega)\gg1$, hence the spectrum function can be written as
\begin{equation}
S(\omega)=\sum_{k}\frac{e^{2}\Xi^{2}_{k}a^{2}k_{B}T}{\hbar^{3}\omega}\delta(\omega-\omega_{k}).
\end{equation}
\end{widetext}

\bibliographystyle{iopart-num}
\bibliography{ChargeNoiseRef}
\end{document}